\documentclass[letterpaper, 10pt, conference]{ieeeconf}  % Comment this line out if you need a4paper
\IEEEoverridecommandlockouts           
\usepackage{graphicx}
\usepackage{epsfig,mathptmx,times,amsmath,amssymb,color,flushend,gensymb,fancyhdr,subfigure,graphicx,cite,soul,framed,lipsum, bm, lipsum}
\usepackage{siunitx}
\usepackage{amsmath, cite}
\let\labelindent\relax
\usepackage{enumitem}
\usepackage{verbatim}
\usepackage{amssymb}
\usepackage[dvipsnames]{xcolor}
\usepackage{mwe}
\usepackage{geometry}
\IEEEoverridecommandlockouts                           

\newtheorem{remark}{Remark}

\title{\LARGE \bf
Concurrent Learning Based Tracking Control of Nonlinear Systems using Gaussian Process
}

\author{Vedant Bhandari and Erkan Kayacan
\thanks{*This work was supported by UQ Early Career Researcher grant (UQECR2057325).}
\thanks{The authors are with the School of Mechanical and Mining Engineering, The University of Queensland, 4072 Brisbane, Australia.
        {\tt\small \{v.bhandari, e.kayacan \}@uq.edu.au}}
}

\begin{document}

\maketitle
\thispagestyle{empty}
\pagestyle{empty}

%%%%%%%%%%%%%%%%%%%%%%%%%%%%%%%%%%%%%%%%%%%%%%%%%%%%%%%%%%%%%%%%%%%%%%%%%%%%%%%%
\begin{abstract}
%This paper demonstrates the applicability of concurrent learning as a tool for parameter estimation in the context of feedback linearization, combined with the use of Gaussian Processes for online learning disturbance rejection. A control law is proposed to use both techniques sequentially and has been shown to be closed-loop stable using the Lyapunov Stability Theorem. Considering systems with no internal dynamics, and dynamics that can be linearly parameterized, simulations have been provided to show minimized tracking error when no model parameter estimates have been initially provided, and in the presence of disturbances introduced once the parameters have converged to their true values, and also in situations where parameters have converged to incorrect values.

This paper demonstrates the applicability of the combination of concurrent learning as a tool for parameter estimation and non-parametric Gaussian Process for online disturbance learning. A control law is developed by using both techniques sequentially in the context of feedback linearization. The concurrent learning algorithm estimates the system parameters of structured uncertainty without requiring persistent excitation, which are used in the design of the feedback linearization law. Then, a non-parametric Gaussian Process learns unstructured uncertainty.  The closed-loop system stability for the nth-order system is proven using the Lyapunov stability theorem.  The simulation results show that the tracking error is minimized (i) when true values of model parameters have not been provided, (ii) in the presence of disturbances introduced once the parameters have converged to their true values and (iii) when system parameters have not converged to their true values in the presence of disturbances.

\end{abstract}

%%%%%%%%%%%%%%%%%%%%%%%%%%%%%%%%%%%%%%%%%%%%%%%%%%%%%%%%%%%%%%%%%%%%%%%%%%%%%%%%
\section{INTRODUCTION}

The performance of feedback linearization depends on the accuracy of the system model, and its parameter estimates in the linearization process and all generalized external forces that change in different environments \cite{westenbroek2020feedback, 8786140, 8930275}. It is often difficult to exactly know the model parameters for any practical system, with uncertainty in measurement always present \cite{OLIVEIRA2020105927, Erkansmlc}. Uncertainty in model parameters leads to a decreasing reference tracking performance \cite{Slotine_Li_1991}. Furthermore, changing operation environments introduces varying disturbances, which cannot always be accounted for in the controller design \cite{Chai}.   

These challenges mentioned above require feedback linearization to be combined with more robust design techniques to account for model uncertainties and adapt to various environments with varying disturbances. The proposed law by \cite{9140024} combines feedback linearization (FBL) control with Gaussian Process (GP) to account for model uncertainty. Using online learning GP, the controller learns the model mismatch between the actual model and the estimated model used for the feedback linearization, reducing the model uncertainty and leading to better reference tracking. However, no attempt is made to estimate the system model's true parameters and the effect of any external disturbances is directly captured by the GP as a model mismatch. This is not ideal as the model mismatch and disturbance cannot be distinguished. A similar proposal by \cite{8264427, BECKERS2019390} also uses GP-based tracking control, where the Gaussian process is used offline to learn the model mismatch and any modelled dynamics together. This presents the same issues as mentioned previously.

%%%%%%%%%%%%%%%%%%%%%%%%%%%%%%%%%%%%%%%%%%%%%%%%%%%%%%%%%%%%%%%%%%%%%%%%%%%%%%%%
% edit this section - Why is parameter estimation important?
%While the other control laws mentioned above result in minimized tracking error, the parameters values are not estimated. Parameter estimation is important and requires attention as it is used in many other control techniques such as Model Predictive Control and allows for the model error to be distinguished from the disturbance.
%%%%%%%%%%%%%%%%%%%%%%%%%%%%%%%%%%%%%%%%%%%%%%%%%%%%%%%%%%%%%%%%%%%%%%%%%%%%%%%%

\begin{figure}
    \centering
    \includegraphics[scale=0.026]{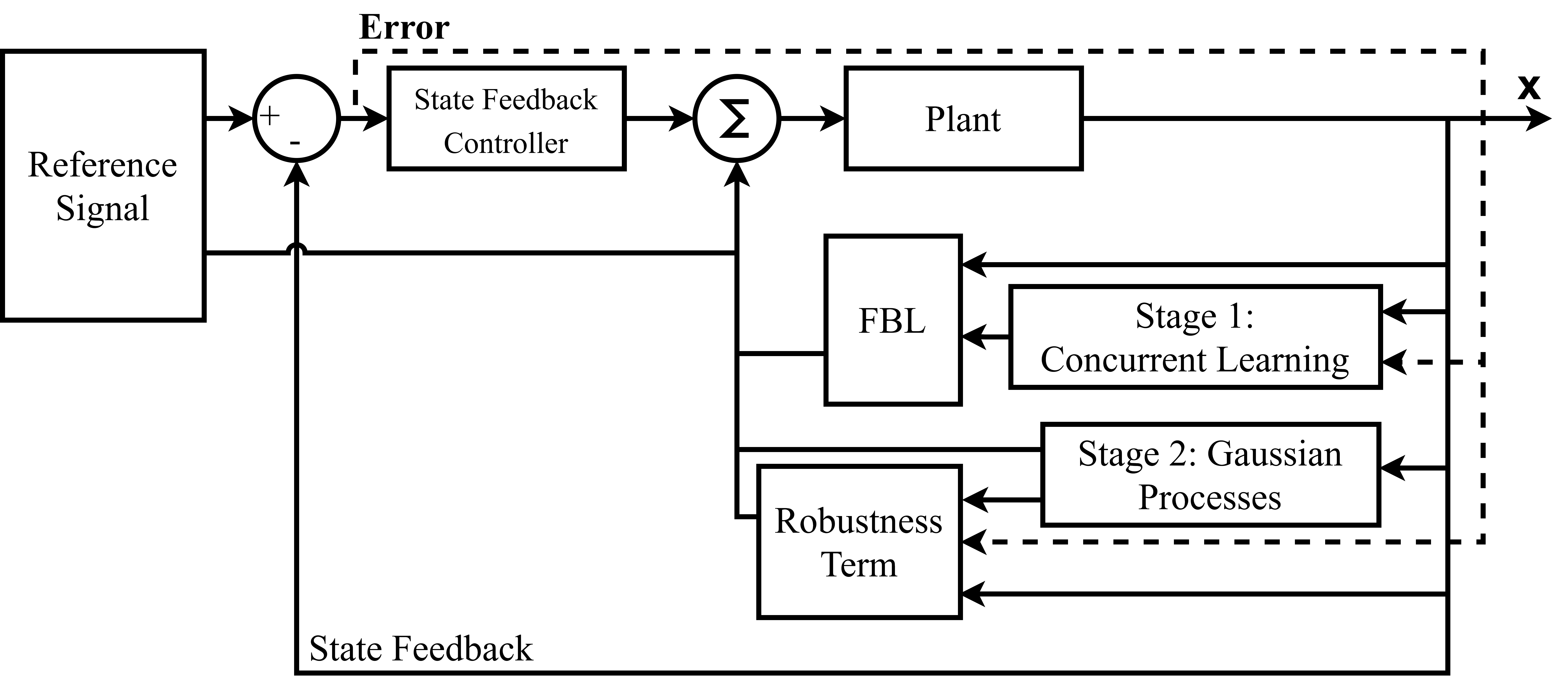}
    \caption{Schematic diagram for the developed control algorithm. Along with the use of the traditional feedback linearization technique (FBL and state-feedback controller blocks), a concurrent learning algorithm is used to estimate the model parameters to be used in the feedback linearization process, and an online non-parametric Gaussian Process and Robustness term are used to account for unknown disturbances.}
    \label{fig:control_diagram}
\end{figure}

%The proposed control law provides a two-stage controller as shown in fig.\ref{fig:control_diagram}. In the first stage, the system model parameters are estimated to provide optimal tracking control with feedback linearization using concurrent learning in a known and controlled environment. Then the learned system model parameters are used with a Gaussian-based online learning approach to operate in environments with unknown disturbances in the second stage. The key benefits of this approach are that the system model parameters can be accurately estimated and the control system has the ability to work in unknown environments.

%Concurrent learning has been used and demonstrated significantly with Model Reference Adaptive Control (MRAC) methods as originally proposed in \cite{5717148}. The proposed approach in this paper demonstrates its applicability in other nonlinear control methods such as feedback linearization – this is unique and has not been found in common control techniques.

Concurrent learning approach was used and demonstrated significantly with model reference adaptive control methods as initially proposed in \cite{5717148}. Recently, it has also been used as an online parameter estimator \cite{8968461}.  It can be used in any other nonlinear control methods. This paper will demonstrate its applicability in the scheme of the FBL method – this is unique and has not been shown in the literature. 

The developed control law in this paper provides a two-stage controller design, as shown in Fig.\ref{fig:control_diagram}. In the first stage, the system model parameters are estimated to provide optimal tracking control with feedback linearization using concurrent learning in a known and controlled environment. The learned model parameters are then used with a GP-based online learning approach to operate in environments with unknown disturbances in the second stage. This approach's key benefits are that the system model parameters can be accurately estimated, which is critical in a model-based controller design process. The unstructured uncertainties can then be captured by a non-parametric GP; thus, the control system can also work in unstructured environments.

%The following contributions are provided in this paper:
%\begin{itemize}
%    \item An approach that uses concurrent learning to accurately estimate system model parameters in the framework of feedback linearization in a known environment.
%    \item A control system that learns the disturbances and model mismatch in the second stage, where Gaussian Processes are used to compensate for the disturbances.
%    \item Showing tracking error convergence with and without exact parameter convergence using Gaussian Processes.
%    \item Stability of the closed-loop system using the Lyapunov Stability Theorem.
%\end{itemize}
%Simulations are provided to to verify the proposed control approach leading to minimized tracking error.

The main contributions of this study are as follow:
\begin{itemize}
    \item An approach that uses concurrent learning to estimate system model parameters in the framework of FBL in a known environment accurately.
    \item A control system that learns the disturbances and model mismatch in the second stage, where a non-parametric GP is used to compensate for the disturbances in unknown, unstructured environments.
    \item The stability of the nth-order closed-loop system is proven using the Lyapunov stability theorem.
    \item Tracking error convergences to zero with and without exact parameter estimation convergence by using GP.
\end{itemize}

This paper is organized as follows: The problem statement is described in Section II. Concurrent learning and Gaussian process are given in Section III. In Section IV, the details of the developed control algorithm is explained, the system stability for an n-th order system is proven. Simulations results have verified the developed control approach leading to minimized tracking error in Section V. Finally, the study is concluded in Section VI.

%%%%%%%%%%%%%%%%%%%%%%%%%%%%%%%%%%%%%%%%%%%%%%%%%%%%%%%%%%%%%%%%%%%%%%%%%%%%%%%%
\section{PROBLEM STATEMENT}
\label{problem_statement}
Single-input single-output (SISO) nonlinear system dynamics that can be linearly parameterized are given by:
%Single-input single-output (SISO) nonlinear system dynamics are %given by:
%\begin{equation}
%    \label{eqn:state_equation}
%    \begin{aligned}
%    &\mathbf{\dot{x}} = \mathbf{f(x)} + \mathbf{g(x)}u + d \\
%    &\mathbf{y} = h\mathbf{(x)}
%    \end{aligned}
%\end{equation}
%with state vector \(\mathbf{x} \in \mathbb{R}^{n}\), \(\mathbf{f}\) and \(\mathbf{g}\) being smooth vector fields, control input \(u\) and an unknown disturbance \(d\) \cite{Slotine_Li_1991}. It is assumed that the system dynamics can be linearly parameterized, such that the unknown system dynamics can be alternatively written by an equivalent form as:
\begin{equation}
    \label{eqn:system_model}
    \mathbf{\dot{x}} = \mathbf{Ax} + \mathbf{b}\big(u+ \mathbf{w}^{*T}{\mathbf{\boldsymbol{\phi}(x)}} + d\big)
\end{equation}
where \(\mathbf{A}\in\mathbb{R}^{n\times n}\) exploits the integrator chain structure, \(\mathbf{b} = [0,0,..., 1]^T \in\mathbb{R}^n \), \(u \in \mathbb{R}\) is the control input, \(\mathbf{x} \in \mathbb{R}^{n}\) is the state vector, the system dynamics are given by \(\mathbf{w}^{*T}\mathbf{\boldsymbol{\phi}(x)}\), and \(d\) is an unknown disturbance. The system model regressor vectors \(\boldsymbol{\phi}(\mathbf{x}) \in \mathbb{R}^m\) are known and \(\mathbf{w}^* \in \mathbb{R}^m\) is the unknown ideal weight vector. Expanding the system equations  \eqref{eqn:system_model}, the state-derivative is given by the last equation:
\begin{equation}
    \label{eqn:W_form}
    \dot{x}_n = {\mathbf{w}^{*T}\boldsymbol{\phi}(\mathbf{x})} + u + d
\end{equation}
where \(\dot{x}_n\) is the state-derivative and \(n\) is the order of the system. 

%The notation is kept consistent with that in \cite{5717148} as the proposed control technique extends from the ideas presented in \cite{5717148}.

%The systems considered in this paper must satisfy the followings:
%\begin{enumerate}
%    \item The system dynamics must be linearly parameterizable.
%    \item The system must have no internal dynamics.
%    \item It is assumed there exists a training environment where the dynamics of the system can be accurately modelled in the regressor vector \(\boldsymbol{\phi}(\mathbf{x})\) before the system is operated in an environment where disturbance \(d\) may exist. This does not necessarily include the weightings of the dynamics, \(\mathbf{w}^*\), but only the dynamics representing the system.
%    \item The state-derivative must be measurable or be estimated.
%\end{enumerate}

In addition to being linearly parameterizable, the system considered in this paper must have no internal dynamics, and the state-derivative in the system model must be measurable or estimated. Moreover, it is assumed there exists a training environment where the dynamics of the system can be accurately modelled in the regressor vector \(\boldsymbol{\phi}(\mathbf{x})\) before the system is operated in an environment where disturbance \(d\) may exist. This does not necessarily include the weightings of the system dynamics, \(\mathbf{w}^*\), but only the dynamics representing the system.

%No. 3 is an important requirement, as if the system cannot be accurately modelled, the parameter estimation using concurrent learning will provide incorrect results, and the Gaussian Process used in the following stage will then compensate for the disturbance, the unknown dynamics and the incorrect parameter estimates. This is undesirable and not the focus of this paper. Ideally, the GP will compensate for the disturbance only.

Feedback linearization algebraically transforms a non-linear system (fully or partially) into a linear one, allowing the use of linear control techniques \cite{Martins, MORENOVALENZUELA2020314}. This is achieved by exact state transformations and feedback control law. 
%Consider the system represented by \eqref{eqn:W_form}, where \(\mathbf{w}^*\) is the ideal system parameter weightings vector. 
Feedback linearization requires a mathematical model of the system and considers a fixed model estimate traditionally \cite{9102437}. The larger the error in the model estimate, the greater the tracking error. Let the estimated weight vector be given by vector \(\mathbf{w}\). To linearize a nonlinear system using feedback linearization, the control input is given by:
\begin{equation}
    \label{eqn:feedback_linearization_only_u}
    u = -\mathbf{w}^T{\boldsymbol{\phi}}(\mathbf{x}) + v
\end{equation}
where \(\mathbf{w}\) is the estimate of the model weights and \(v\) is a selected linear controller. This results in the following system:
\begin{equation}
    \label{eqn:linear_controller}
    \dot{x}_n = -\Tilde{\mathbf{w}}\boldsymbol{\phi}(\mathbf{x}) + d + v
\end{equation}
where \(\mathbf{\Tilde{w} = w - w^*}\) is the modelling error. Note that if \(\mathbf{w \rightarrow w^*}\), then \(\mathbf{\Tilde{w}} = 0\) and also if \(d = 0\), then a linear system is obtained as
\begin{equation}
    \label{eqn:linear_system}
    \dot{x}_n = v    
\end{equation}

Considering the closed-loop dynamics presented by \eqref{eqn:linear_system}, the linear controller \(v\) is chosen such that the closed-loop system is stable. Let the tracking error be defined as :
\begin{equation}
    \mathbf{e = x_{ref} - x}
\end{equation}
where \(\mathbf{e}\) is the vector of the error states and \(\mathbf{x_{ref}}\) is the vector of the state references where
\begin{equation}
    \mathbf{e}=\begin{bmatrix} e_1 & e_2 & \cdots & e_n \end{bmatrix}^T, \enspace e_i=x_{i_{\mathbf{ref}}} - x_i
\end{equation}
The linear controller, i.e., \(v\), is designed such that:
\begin{equation}\label{eqn:linear_controlv}
    v =\dot{x}_{n_{\mathbf{ref}}} + k_1 e_1 + k_2 e_2 + \cdots + k_ne_n
\end{equation}
Substituting the linear controller \(v\) \eqref{eqn:linear_controlv} in \eqref{eqn:linear_system} gives: 
\begin{equation}
    \label{eqn:error_polynomial}
    0 = \dot{e}_n + k_n e_n + \cdots + k_2 e_2 + k_1 e_1
\end{equation}
The values of the polynomial coefficients \(k_1, k_2, \cdots, k_n \geq 0\) are such that the system poles are on the left half of the plane, leading to exponentially stable dynamics \cite{Slotine_Li_1991}. This demonstrates that the control input given by \eqref{eqn:feedback_linearization_only_u} results in stable closed-loop dynamics when the system is perfectly modelled and in the absence of any disturbances \cite{8944010}.

However, in the common case that \(\mathbf{w \neq w^*}\), and in the presence of a disturbance, \eqref{eqn:error_polynomial} becomes:
\begin{equation}
    \label{eqn:error_polynomial_errors}
     \mathbf{\Tilde{w}}\boldsymbol{\phi}(\mathbf{x}) - d = \dot{e}_n + k_n e_n + \cdots + k_2 e_2 + k_1 e_1 
\end{equation}
Eq. \eqref{eqn:error_polynomial_errors} shows that the tracking error cannot converge to zero due to the presence of modelling error and disturbance under the  traditional feedback linearization control law proposed in \eqref{eqn:feedback_linearization_only_u}. As the control law is deemed insufficient in this case, it needs to be redesigned to account for system modelling errors and disturbances. 

%Then, the closed-loop system is not exponentially stable as before. \eqref{eqn:error_polynomial_errors} does not guarantee error convergence to zero due to the presence of modelling error and disturbance. The traditional feedback linearization control input given by \eqref{eqn:feedback_linearization_only_u}, is deemed insufficient in this case and needs to be redesigned to account for system modelling errors and disturbances. 

The goal of the developed control law in this paper is to design the control input \(u\) such that the tracking error is minimized and converges to zero in finite-time in the presence of modelling error and disturbance.
%, the system model parameters can be accurately estimated, and the closed-loop system is stable when:
%\begin{itemize}
%    \item Scenario 1: The model parameters are incorrectly estimated and then exposed to uncertain environments.
%    \item Scenario 2: The model parameters are correctly estimated and then exposed to uncertain environments.
%\end{itemize}
This goal is to be achieved using concurrent learning to accurately estimate the system parameters to minimize the modelling error and Gaussian Process to learn the disturbance and cancel its effects by compensating in the control input.

%%%%%%%%%%%%%%%%%%%%%%%%%%%%%%%%%%%%%%%%%%%%%%%%%%%%%%%%%%%%%%%%%%%%%%%%%%%%%%%%
\section{CONCURRENT LEARNING AND GAUSSIAN PROCESS}
%%%%%%%%%%%%%%%%%%%%%%%%%%%%%%%%%%%%%%%%%%%%%%%%%%%%%%%%%%%%%%%%%%%%%%%%%%%%%%%%
%%%%%%%%%%%%%%%%%%%%%%%%%%%%%%%%%%%%%%%%%%%%%%%%%%%%%%%%%%%%%%%%%%%%%%%%%%%%%%%%
\subsection{Concurrent Learning}
\label{concurrent_learning}
%%%%%%%%%%%%%%%%%%%%%%%%%%%%%%%%%%%%%%%%%%%%%%%%%%%%%%%%%%%%%%%%%%%%%%%%%%%%%%%%
%Concurrent learning allows for system model parameter estimation instead of providing a user-selected model parameter estimate, which is difficult to obtain accurately. Traditionally, the constant model parameter estimate in the feedback linearization control method leads to a constant error in the linearization process. It would be ideal to update the model parameter weights using state feedback until the model parameters converge to their true values. Using the concept of concurrent learning introduced by \cite{5717148}, the model parameters are updated until the weight vector \(\mathbf{w\rightarrow w^*}\) by learning the system. Concurrent learning uses a set of recorded data to converge parameters to their true values and not only the current value, as visible in the second half of \eqref{eqn:concurrent_learning_law}.

Traditionally, an inaccurate model parameter estimate in the feedback linearization control method leads to an error in the linearization process, and it is not easy to obtain the model parameter estimate accurately. It would be ideal to update the model parameter weights using state feedback until the model parameters converge to their true values. Concurrent learning allows for model parameter estimation without requiring persistent excitation instead of providing a user-selected model parameter estimate. It uses the current error and a set of recorded data to converge parameter estimates to their true values, as visible in the second half of \eqref{eqn:concurrent_learning_law}.

The concurrent learning method is initially introduced in the context of model reference adaptive control. The same concept is integrated with the feedback linearization framework in this paper. The following weight update law is used:
\begin{equation}
    \label{eqn:concurrent_learning_law}
    \dot{\mathbf{w}} = -\Gamma_W\boldsymbol{\phi}(\mathbf{x})\mathbf{e}^T\mathbf{Pb} - \sum_{j=1}^{p} \Gamma_W\boldsymbol{\phi}(\mathbf{x})\epsilon_j
\end{equation}
with the conditions and parameters detailed in \cite{5717148}. The algorithm to store data points is given in \cite{8968461}. The Lyapunov equation is written as follows:
%\(\mathbf{P}\in \mathbb{R}^{n\times n}\) is the positive definite solution to the Lyapunov equation \cite{Boley1994NumericalMF}:
\begin{equation}
    \mathbf{(A-bk)}^T\mathbf{P} + \mathbf{P(A-bk)} + \mathbf{Q} = 0
\end{equation}
where \(\mathbf{(A-bk)}\) is introduced in Section \ref{tracking_error}, \(\mathbf{Q}\in \mathbb{R}^{n\times n}\) is any positive definite matrix and \(\mathbf{P}\in \mathbb{R}^{n\times n}\) is the positive definite solution to the Lyapunov equation \cite{Boley1994NumericalMF}. It was shown the weight update law to be exponentially stable within the model reference adaptive control context \cite{5717148}. The error dynamics of the proposed closed-loop system in the context of feedback linearization is formulated in Section \ref{tracking_error} and is shown to be in the form as in \cite{5717148}, meaning the stability results apply directly to the proposed closed-loop system as well. The stability results for the parameter convergence are not repeated here and can be found in \cite{5717148, 7858671}.

%%%%%%%%%%%%%%%%%%%%%%%%%%%%%%%%%%%%%%%%%%%%%%%%%%%%%%%%%%%%%%%%%%%%%%%%%%%%%%%%
\subsection{Gaussian Process}
\label{Gaussian_Processes}
Gaussian Processes are used in the control law to compensate for disturbances when systems exposed to an unknown environment \cite{BRADFORD2020106844, BECKERS2019390}. The following part of this subsection outlines the key concepts used in the controller design in this paper.

Gaussian Process (GP) Regression allows for a nonlinear mapping from a set of input values \(X=\)\{\(x_1,x_2,\cdots,x_n\)\} to output values \(Y=\)\{\(y_1,\cdots,y_n\)\} \cite{9285210}. The GP is provided with a set of training data \(D=\)\{\(X,Y\)\}, from which a nonlinear map is generated, so an unseen input \(X\) returns a predicted output \(Y\) \cite{Rasmussen_Williams_2006}. This makes GP a powerful prediction tool and useful addition to the control law discussed in Section \ref{control_law} when dealing with disturbances. The idea of Bayesian inference is used to perform regression on a set of training data, in which the current hypothesis is updated based on new information. If the greater number of training data is provided around a point to be estimated, the better GP prediction is obtained.

Two priors are required to define the Gaussian distribution along with the training data: a mean function \(\mu\), which describes the expected value of the distribution, generally set to zero, and a covariance matrix or kernel, \(\sum\), which provides a measure of similarity between two points and their values, describing the shape of the distribution. These two priors can significantly impact the nonlinear mapping generated by the GP and must be selected carefully. Each kernel has various hyperparameters that are also user-selected and significantly impact the overall input-output mapping. However, these parameters can be optimized by solving the maximum log-likelihood problem \cite{Rasmussen_Williams_2006}.

Mathematically, consider the joint Gaussian distribution of the training outputs, \(Y\) corresponding to inputs \(X\), and the test outputs, \(Y_*\) corresponding to test inputs \(X_*\) \cite{Rasmussen_Williams_2006}:
\begin{equation}
    \begin{bmatrix} Y \\ Y_* \end{bmatrix} \sim \mathcal{N} \left(0, \begin{bmatrix}
    K(X,X) & K(X,X_*) \\ K(X_*, X) & K(X_*, X_*) \end{bmatrix}\right)
\end{equation}
In the controller, the test inputs, \(X_*\), are the time-dependent system states. For \(n\) training points and \(n_*\) test points, \(K(X,X_*)\) represents the \(n \times n_*\) matrix of covariances evaluated at all pairs of training and test points. The same applies to the other entries of the covariance function. The squared exponential covariance function is used in this paper and is given by \cite{Duvenaud_2014}:
\begin{equation}
    k_{SE}(X,X_*) = \sigma^2_f\exp{\left(-\frac{(X-X_*)^2}{2l^2}\right)}
\end{equation}

\begin{figure}
    \centering
    \includegraphics[scale=0.04]{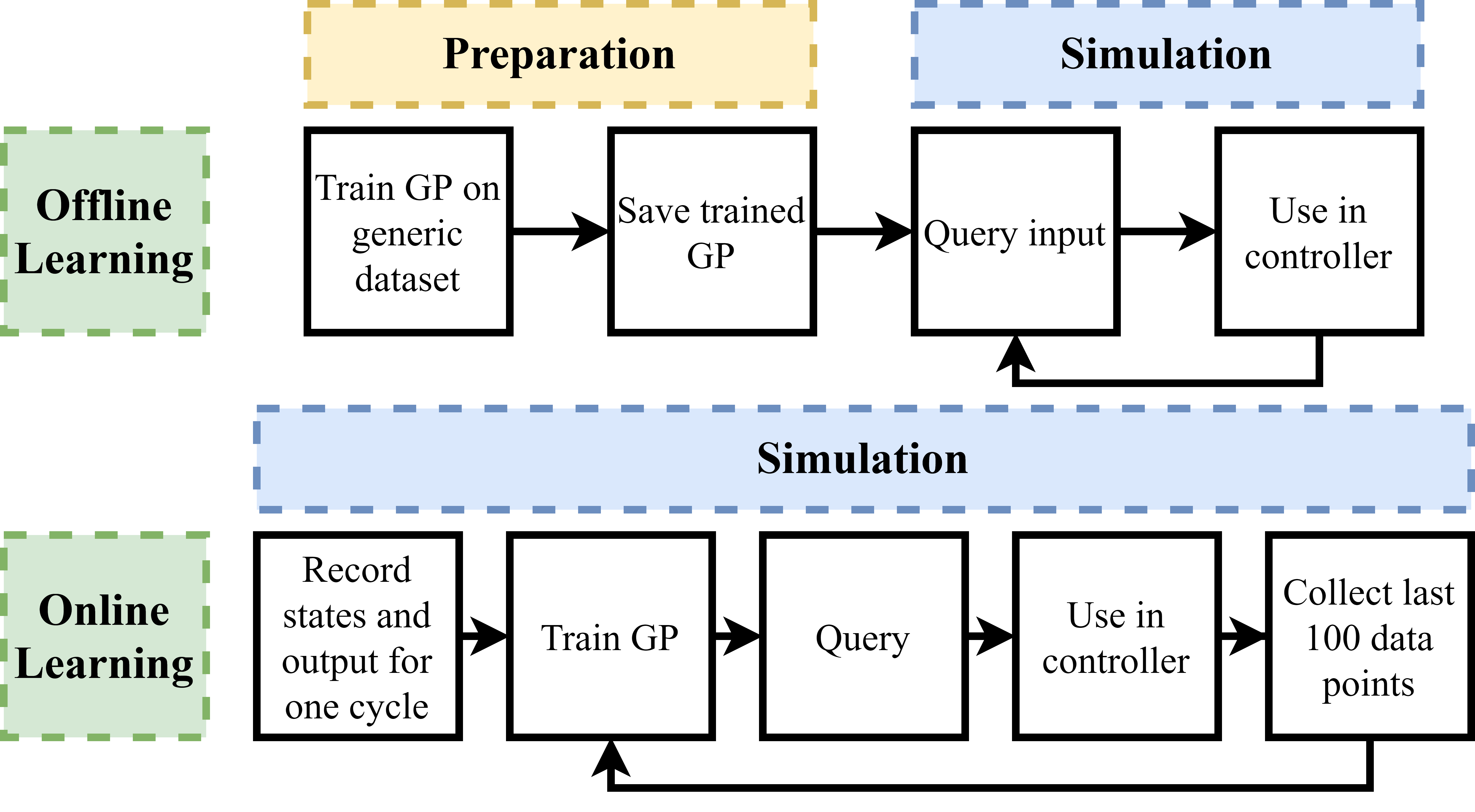}
    \caption{Gaussian Process. Offline vs Online Gaussian Process learning.}
    \label{fig:GP_learning}
\end{figure}

Conditioning the joint prior distribution on the training data, \(D\), the mean and variance at query points \(X_*\) is given by \cite{Rasmussen_Williams_2006}:
\begin{equation}
    \label{eqn:Gaussian_mean}
    \mu(X_*) = K(X_*,X)K(X,X)^{-1}Y
\end{equation}
\begin{equation}
    \resizebox{.89\hsize}{!}{$\sigma^2(X_*) = K(X_*,X_*)-K(X_*,X)K(X,X)^{-1}K(X,X_*)$}
\end{equation}
In the context of this paper, a GP is trained with \(n\) data pairs, with the system states, \(\textbf{x}\) as inputs, and the control input mismatch \((u - \hat{u})\), as the output, representing the torque compensation for the disturbance. This can be performed using two methods; online or offline learning displayed in Fig. \ref{fig:GP_learning}.

In offline learning, the GP is trained across a set of generic inputs, and the pre-recorded GP is used to predict the compensation for the disturbance in  simulation. This differs from online learning in the sense that the GP is not updated during the simulation. In online learning, the GP model is continuously updated throughout tracking the reference signal. This allows the GP to be trained to meet the experimental conditions and update according to any model mismatch during the simulation. Using an online GP requires greater processing power as the prediction hypothesis is updated throughout the simulation compared to only obtaining a prediction from the GP. 

%%%%%%%%%%%%%%%%%%%%%%%%%%%%%%%%%%%%%%%%%%%%%%%%%%%%%%%%%%%%%%%%%%%%%%%%%%%%%%%%
\section{METHODOLOGY}
%%%%%%%%%%%%%%%%%%%%%%%%%%%%%%%%%%%%%%%%%%%%%%%%%%%%%%%%%%%%%%%%%%%%%%%%%%%%%%%%
%\subsection{System Model}
%\label{system_model_form}
%Consider a nonlinear system whose dynamics can be linearly parameterized. The system is given by:
%\begin{equation}
%    \label{eqn:system_model}
%    \mathbf{\dot{x}} = \mathbf{Ax} + \mathbf{B}(\mathbf{u}+ %W^{*T}\Phi({x}) + d)
%\end{equation}
%where \(\mathbf{A}\in\mathbf{R}^{n\times n}\) exploits the integrator chain structure and \(\mathbf{B}\in\mathbf{R}^n, B = [0,0,..., 1]^T\), \(u(t)\) is the control signal, the system dynamics are given by \(W^{*T}\Phi(\mathbf{x})\), and \(d\) is a disturbance unknown to the control system.
%%%%%%%%%%%%%%%%%%%%%%%%%%%%%%%%%%%%%%%%%%%%%%%%%%%%%%%%%%%%%%%%%%%%%%%%%%%%%%%%
%The system model presented in Section \ref{problem_statement} is used in the following.
\subsection{Reference Model}
\label{reference_model}
In matrix form, the reference system can be represented as:
\begin{equation}
    \label{eqn:reference_model}
    \mathbf{\dot{x}_{ref}} = \mathbf{Ax_{ref}} + \mathbf{b}u
\end{equation}
where \(\mathbf{A}\in\mathbb{R}^{n\times n}\) exploits the integrator chain structure, \(\mathbf{x_{ref}}=[x_{1_{\mathbf{ref}}}, x_{2_{\mathbf{ref}}}, \cdots, x_{{n-1}_{\mathbf{ref}}}]\) is the reference output, \(\mathbf{b}\in\mathbb{R}^n, \mathbf{b} = [0,0,..., 1]^T\) as before and \(u\) is the reference input equal to the reference output, \(u = \dot{x}_{n_{\mathbf{ref}}}\). The reference signal is assumed to be continuously differentiable.
%%%%%%%%%%%%%%%%%%%%%%%%%%%%%%%%%%%%%%%%%%%%%%%%%%%%%%%%%%%%%%%%%%%%%%%%%%%%%%%%
\subsection{Control Law}
\label{control_law}
Let the control law be defined as:
\begin{equation}
    \label{eqn:control_law_u}
    u = u_{FBL} + u_{SFB} + u_{ref} - u_{GP} - u_{rob}
\end{equation}
where \(u_{FBL}\) is the feedback linearization law to linearize the system, \(u_{SFB}\) is the chosen linear controller, \(u_{ref}\) is the reference signal required to cancel the input from the reference model, \(u_{GP}\) is the Gaussian process-based estimate of the disturbance, and \(u_{rob}\) is robustness term with its formulation detailed in Section \ref{system_stability} to ensure closed-loop system stability. They are formulated as follows:
\begin{eqnarray}
    u_{FBL} &=& -\mathbf{w}^T\boldsymbol{\phi}(\mathbf{x})  \\
   u_{SFB} &=& \mathbf{ke}, \enspace \mathbf{k} = [k_1, k_2, \cdots, k_n] \\
   u_{ref} &=& \dot{x}_{n_{\mathbf{ref}}} \\
   u_{GP} &=& \mu(\mathbf{x^*}) = u(\mathbf{x^*}) - \hat{u}(\mathbf{x^*}) \\
   \label{eqn:u_rob_term}
   u_{rob} &=& -m  \frac{\mathbf{b}^T\mathbf{Pe}}{\lVert \mathbf{b}^T\mathbf{Pe} \rVert}
\end{eqnarray}
%\begin{equation}
%    \label{eqn:FBL}
%    u_{FBL} = -\mathbf{w}^T\boldsymbol{\phi}(\mathbf{x})
%\end{equation}
%\begin{equation}
%    \label{eqn:PD}
%    u_{PD} = \mathbf{ke}, \enspace \mathbf{k} = [k_1, k_2, \cdots, k_n]
%\end{equation}
%\begin{equation}
%    \label{eqn:ref}
%    u_{ref} = \dot{x}_{n_{\mathbf{ref}}}
%\end{equation}
%\begin{equation}
%    \label{eqn:GP}
%    u_{GP} = \mu(\mathbf{x^*}) = u(\mathbf{x^*}) - \hat{u}(\mathbf{x^*})
%\end{equation}
%\begin{equation}
    %u_{rob} = -(1+\rho)\lvert \mathbf{\Tilde{w}}\boldsymbol{\phi}(\mathbf{x}) - d + \mu(\mathbf{x}) \rvert  \frac{\mathbf{b}^T\mathbf{Pe}}{\lVert \mathbf{b}^T\mathbf{Pe} \rVert}
%    u_{rob} = -m  \frac{\mathbf{b}^T\mathbf{Pe}}{\lVert \mathbf{b}^T\mathbf{Pe} \rVert}
%\end{equation}
\(u_{GP}\) is determined at a query state \(\mathbf{x^*}\). The estimate is given by learning the difference in the actual control input, i.e., \(u\), and the control input estimated from the known system model, i.e., \(\hat{u}\). If the model parameters converge to their true value during the learning stage, only the disturbance is learnt by the Gaussian Process. The plant model is given by:
\begin{equation}
    u = \dot{x}_n - \mathbf{w}^{*T}\boldsymbol{\phi}(\mathbf{x}) - d
\end{equation}
The estimated model is given by:
\begin{equation}
    \hat{u} = \dot{x}_n - \mathbf{w}^{T}\boldsymbol{\phi}(\mathbf{x})
\end{equation}
Using the same state feedback, \(\mathbf{x}\), \(u-\hat{u} = - d\) if \(\mathbf{w\rightarrow w^*}\). If the parameters do not converge, the GP learns the disturbance and the model convergence error.
%%%%%%%%%%%%%%%%%%%%%%%%%%%%%%%%%%%%%%%%%%%%%%%%%%%%%%%%%%%%%%%%%%%%%%%%%%%%%%%%
\subsection{Tracking Error}
\label{tracking_error}
The error dynamics are  given by:
%\begin{equation}
%    \label{eqn:tracking_error}
%    \mathbf{e} = \mathbf{x_{ref}} - \mathbf{x}
%\end{equation}
\begin{equation}
    \label{eqn:tracking_error_dynamics}
    \dot{\mathbf{e}} = \mathbf{\dot{x}_{ref}} - \dot{\mathbf{x}}
\end{equation}
Substituting the reference and system model in (\ref{eqn:tracking_error_dynamics}), the error dynamics can be written considering \eqref{eqn:reference_model} as:
\begin{equation}
    \label{eqn:tracking_error_dynamics_substitued}
    \dot{\mathbf{e}} = \mathbf{Ax_{ref}} + \mathbf{b}\dot{x}_{n_{ref}} - \Big(\mathbf{A}\mathbf{x} + \mathbf{b}\big(u + \mathbf{w}^{*T}\boldsymbol{\phi}(\mathbf{x})\big)\Big)
\end{equation}
Substituting in for the control law (\ref{eqn:control_law_u}) and letting \(\mathbf{\Tilde{w} = w - w^*}\) be the model parameter estimation error, the error dynamics can be simplified to:
\begin{equation}
    \label{eqn:tracking_error_dynamics_simplified}
    \dot{\mathbf{e}} = (\mathbf{A-bk})\mathbf{e} + \mathbf{b}\big(\mathbf{\Tilde{w}}\boldsymbol{\phi}(\mathbf{x})-d+\mu(\mathbf{x^*}) + u_{rob}\big)
\end{equation}
%%%%%%%%%%%%%%%%%%%%%%%%%%%%%%%%%%%%%%%%%%%%%%%%%%%%%%%%%%%%%%%%%%%%%%%%%%%%%%%%
%\section{System Stability}
%The error dynamics have been defined in \eqref{eqn:tracking_error_dynamics_simplified}. Two cases are possible:
%\begin{enumerate}
%    \item The concurrent learning accurately parametrizes the system model, so \(\mathbf{\Tilde{w}} = 0\).
%    \item The concurrent learning has an error, so \(\mathbf{\Tilde{w}}\neq 0\).
%\end{enumerate}
%Case 2 captures both scenarios and will be shown to have stable closed loop tracking error dynamics, satisfying both cases. In the case that \(\mathbf{\Tilde{w}} \neq 0\), the Gaussian process will also learn the mismatch. 
%%%%%%%%%%%%%%%%%%%%%%%%%%%%%%%%%%%%%%%%%%%%%%%%%%%%%%%%%%%%%%%%%%%%%%%%%%%%%%%%
\subsection{System Stability}
\label{system_stability}
The closed-loop system is shown to be stable using the Lyapunov stability theorem in the possibility of concurrent learning not learning the system parameters correctly as well. An approach similar to \cite{9140024} is used below to show the closed-loop system is stable. The common quadratic Lyapunov function is proposed below:
\begin{equation}
    \label{eqn:Lyapunov_function}
    V = \mathbf{e}^T\mathbf{Pe}
\end{equation}
where \(\mathbf{e}\) is the tracking error vector and \(\mathbf{P}\) is the positive definite matrix that solves the algebraic Ricatti equation:
\begin{equation}
    \label{eqn:Algebraic_Ricatti_Equation}
    \mathbf{(A-bk)}^T\mathbf{P+P(A-bk)+\Tilde{S}} = 0
\end{equation}
where \(\mathbf{\Tilde{S} = Q+k^TRk > 0}\) with \(\mathbf{Q,R} > 0\). The derivative of the Lyapunov function is given by:
\begin{equation}
    \label{eqn:Lyapunov_function_derivative}
    \dot{V} = \mathbf{e}^T\mathbf{P\dot{e}}
\end{equation}
The time-derivative of Lyapunov function is equivalently using the algebraic Ricatti equation and by inserting \eqref{eqn:tracking_error_dynamics_simplified} into the equation above:
\begin{equation}
    \label{eqn:Ricatti_equation_derivative}
    \dot{V} = \mathbf{-e}^T\mathbf{\Tilde{S}e} + 2\mathbf{(b}^T\mathbf{Pe)}^T\big(\mathbf{\Tilde{w}}\boldsymbol{\phi}(\mathbf{x})-d+\mu(\mathbf{x^*}) + u_{rob}\big)
\end{equation}
 The robustness term \(u_{rob}\) is now designed such that the system can be shown to be stable. Using \(u_{rob}\) defined in \eqref{eqn:u_rob_term} and substituting into \eqref{eqn:Ricatti_equation_derivative}:

\begin{align}
     \label{eqn:Ricatti_equation_derivative_with_u_rob}
     \dot{V} = &\mathbf{-e}^T\mathbf{\Tilde{S}e} + 2\mathbf{(b}^T\mathbf{Pe)}^T  \nonumber \\
     & \Big(\mathbf{\Tilde{w}}\boldsymbol{\phi}(\mathbf{x})-d+\mu(\mathbf{x^*}) -m\frac{\mathbf{b}^T\mathbf{Pe}}{\lVert \mathbf{b}^T\mathbf{Pe} \rVert}\Big)
\end{align}

Using the Lyapunov theorem, \(\dot{V}\) needs to be shown to be strictly negative for the system to be closed-loop stable. The first term \(\mathbf{-e^T\Tilde{S}e < 0}\) is strictly negative. The second term also needs to be shown to be strictly negative to show system stability. Let the second term be represented by \(V_2\). Using the Cauchy-Schwarz inequality, \(V_2\) can be rewritten as:
\begin{equation}
    \label{eqn:Cauchy-Schwarz_Inequality}
    V_2 \leq 2 \lVert \mathbf{b}^T\mathbf{Pe} \rVert \big(-m + \lvert \mathbf{\Tilde{w}}\boldsymbol{\phi}(\mathbf{x})-d+\mu(\mathbf{x^*}) \rvert\big)
\end{equation}
\(V_2\) is negative if and only if:
\begin{equation}
    % \label{}
    m >  \lvert \mathbf{\Tilde{w}}\boldsymbol{\phi}(\mathbf{x})-d+\mu(\mathbf{x^*}) \rvert
\end{equation}
%To guarantee \(V_2\) to be negative, \(m\) is designed such that:
%\begin{equation}
%    \label{eqn:u_rob_m}
%    m = (1+\rho) \lvert \mathbf{\Tilde{w}}\boldsymbol{\phi}(\mathbf{x})-d+\mu(\mathbf{x^*}) \rvert
%\end{equation}
%where \(\rho > 0\) is a user-selected value, e.g. \(\rho = 0.01\). Substituting this for m:
%\begin{align}
% V_2 \leq & \enspace 2\lVert \mathbf{b}^T\mathbf{Pe} \rVert \big[-(1+\rho) \lvert \mathbf{\Tilde{w}}\boldsymbol{\phi}(\mathbf{x})-d+\mu(\mathbf{x^*}) \rvert \nonumber \\
%   & + \lvert \mathbf{\Tilde{w}}\boldsymbol{\phi}(\mathbf{x})-d+\mu(\mathbf{x^*}) \rvert \big]
%\end{align}

%Because \(2\lVert \mathbf{b}^T\mathbf{Pe} \rVert > 0\), and 
%\begin{equation}
%    %\label{eqn:}
%    (1+\rho) \lvert \mathbf{\Tilde{w}}\boldsymbol{\phi}(\mathbf{x})-d+\mu(\mathbf{x^*}) \rvert > \lvert \mathbf{\Tilde{w}}\boldsymbol{\phi}(\mathbf{x})-d+\mu(\mathbf{x^*}) \rvert
%\end{equation}
%\(V_2\) is shown to be negative, or equal to zero when there is no disturbance. 

Therefore, as \(V_1\) is strictly negative, and \(V_2\) is negative in the presence of a disturbance or no disturbance, \(\dot{V} = V_1 + V_2 < 0\) and the closed-loop system is stable. 

\begin{remark}
The term above \(\lvert \mathbf{\Tilde{w}}\boldsymbol{\phi}(\mathbf{x})-d+\mu(\mathbf{x^*}) \rvert\) is equal to $ \lvert \dot{x}_{n_{measured}} - \dot{x}_n \rvert $.
\end{remark}
%%%%%%%%%%%%%%%%%%%%%%%%%%%%%%%%%%%%%%%%%%%%%%%%%%%%%%%%%%%%%%%%%%%%%%%%%%%%%%%%
%\subsection{Control Law}
%The control law is given by (\ref{eqn:control_law_u}) with \(u_{FBL}\), \(u_{PD}\), \(u_{ref}\), \(u_{GP}\) are given by (\ref{eqn:FBL}), (\ref{eqn:PD}), (\ref{eqn:ref}), (\ref{eqn:GP}) respectively, and \(u_{rob}\) is given by (\ref{eqn:u_rob}) with \(m\) defined in (\ref{eqn:u_rob_m}).
%%%%%%%%%%%%%%%%%%%%%%%%%%%%%%%%%%%%%%%%%%%%%%%%%%%%%%%%%%%%%%%%%%%%%%%%%%%%%%%%
\section{Simulation}
\label{simulation}
The simulation results for the proposed control law applied to the system presented in \cite{5717148} are presented in this section. The following cases are simulated for the system and referred to below:

\renewcommand{\labelenumi}{Case \alph*)}
\begin{enumerate}[leftmargin=1.5cm]
%[label=\alph*)]
    \item FBL without any learning, with model mismatch  \(\mathbf{\Tilde{w}} \neq 0\) in the absence of disturbance, \(d=0\).
    \item FBL with only concurrent learning for parameter estimation without model mismatch  \(\mathbf{\Tilde{w}} = 0\) in the absence of disturbance, \(d=0\).
    \item FBL with only concurrent learning for parameter estimation without model mismatch  \(\mathbf{\Tilde{w}} = 0\) in the presence of disturbance, \(d\neq 0\).
    \item FBL with concurrent learning for parameter estimation with model mismatch \(\mathbf{\Tilde{w}} \neq 0\) and Gaussian process-based online learning for disturbance in the presence of disturbance, \(d\neq0\). 
    \item FBL with concurrent learning for parameter estimation without model mismatch \(\mathbf{\Tilde{w}} = 0\) and Gaussian process-based online learning for disturbance in the presence of disturbance, \(d\neq0\). 
\end{enumerate}

%The two scenarios outlined in Section \ref{problem_statement} are highlighted in cases d and e, respectively. 
% The aims of this paper are aligned with case (e), in which model parameters are accurately estimated.
All simulations use gains \(k_1=k_2=20\), \(\rho=0.01\) and a reference tracking of \(x_{ref}=A \sin{\omega t}\) with \(A=0.5\) and \(\omega = 1 \enspace rad/s\). All initial conditions are zero, i.e., \(\textbf{x}(0)=\textbf{0}\). A learning rate of \(\Gamma_W = 3\) is used for the concurrent parameter estimation. For the GP, the squared exponential kernel is used as each new training point reduces the posterior variance \cite{Steinwart_Christmann_2008}. The hyperparameters are optimized using the maximum log likelihood function and all GP processing is handled by using the \textit{fitrgp} function in MATLAB.

The online GP implementation has a ten-second learning period where the model is initially learnt, and then the model is updated with every previous 100 data points throughout the simulation. Since the disturbance is unknown to a control system in practice, it is not feasible to use offline GP learning unless the same simulation was run under the same circumstances, the model was trained and then deployed again. Using an online GP allows the learning and training to happen simultaneously and is used below for all simulations.
%%%%%%%%%%%%%%%%%%%%%%%%%%%%%%%%%%%%%%%%%%%%%%%%%%%%%%%%%%%%%%%%%%%%%%%%%%%%%%%%
\subsection{Systems}
\label{simualtion_systems}
The system used for simulations is the same as that presented in \cite{5717148}, and is given by:
\begin{equation}
    \label{eqn:system_1}
    \ddot{\theta} = u + \sin{\theta} - \lvert\dot{\theta}\rvert \theta + 0.5\exp{(\theta\dot{\theta})}
\end{equation}
Rewriting in the form of \eqref{eqn:W_form}:
\begin{equation}
    \label{eqn:system_1_system_format}
    \ddot{\theta} =  \mathbf{w}^{*T}\boldsymbol{\phi}(\boldsymbol{\theta}) + u + d
\end{equation}
where the ideal weights, \(\mathbf{w}^* = [ 1, -1, 0.5]^T\), the regressor vector values are \(\boldsymbol{\phi}(\boldsymbol{\theta}) = [ \sin{\theta}, \lvert\dot{\theta}\rvert\theta, \exp{(\theta\dot{\theta})}]^T\), and \(u\) is the control input. For Case a), the weight estimate is given by  \(\mathbf{w} = [0.5, -1.3, 0.75 ]^T\). The added external disturbance is given by:
\begin{equation}
d = \begin{cases}
    0 & 0 < t < 10 \\
    \cos{\theta} + \dot{\theta} & 10 \leq t \leq 30 
\end{cases}
\end{equation}
The simulation has three stages. In the first ten seconds, there is no disturbance and concurrent learning is used to estimate the model parameters. From 10 to 20 seconds, the disturbance is introduced and data is collected to train the GP. After 20 seconds, the GP begins compensating for the disturbance and the GP is continuously updated. 
%%%%%%%%%%%%%%%%%%%%%%%%%%%%%%%%%%%%%%%%%%%%%%%%%%%%%%%%%%%%%%%%%%%%%%%%%%%%%%%%
\subsection{Simulation Results}

\begin{figure*}[t!]
\centering
\subfigure[Average tracking error (\%).]{
\includegraphics[width=0.64\columnwidth]{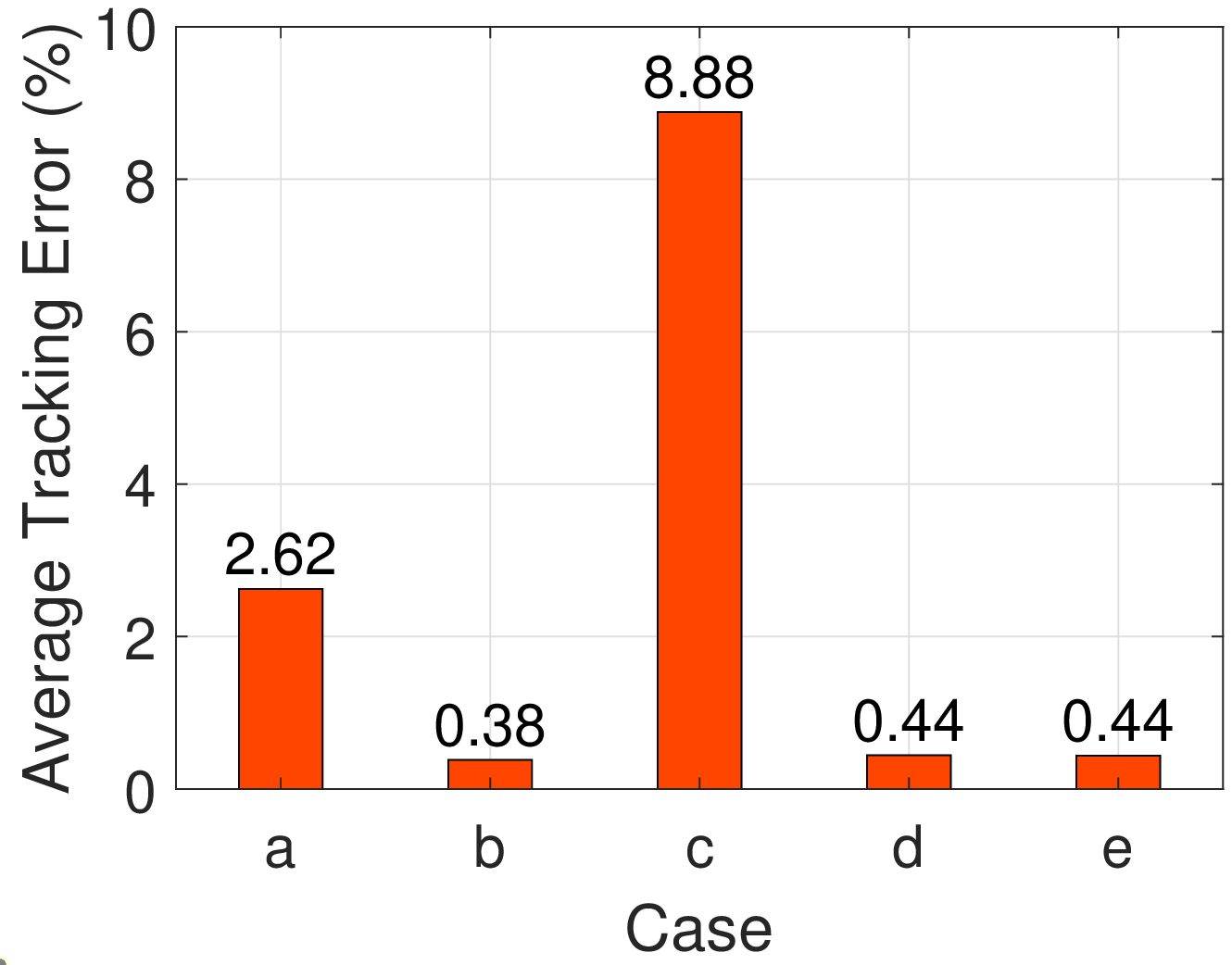}
\label{fig:combined_average_tracking_error}
} 
\subfigure[Reference tracking Case e).]{
\includegraphics[width=0.64\columnwidth]{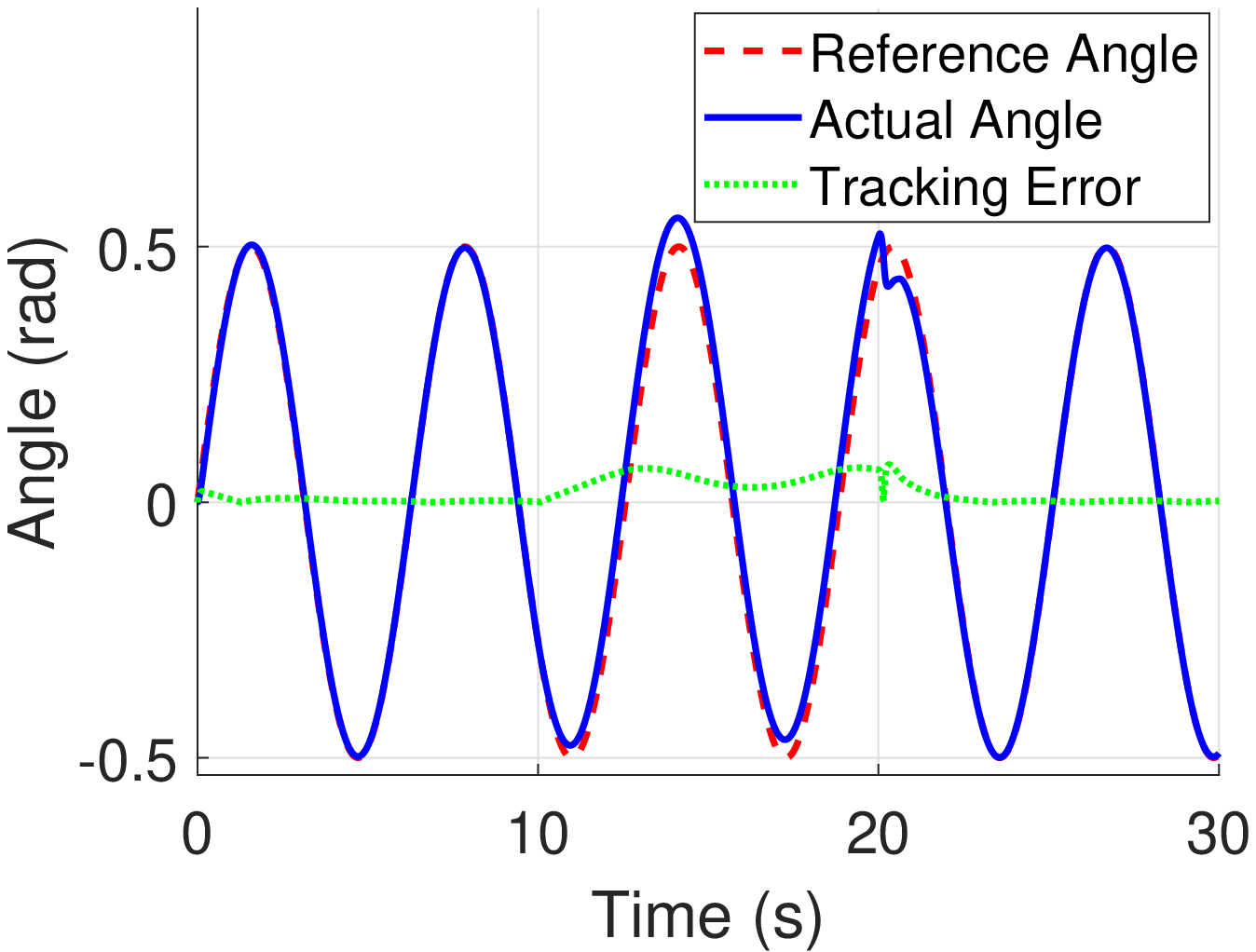}
\label{fig:correct_model_reference_signal_system}
}  
\subfigure[Reference tracking Case d).]{
\includegraphics[width=0.64\columnwidth]{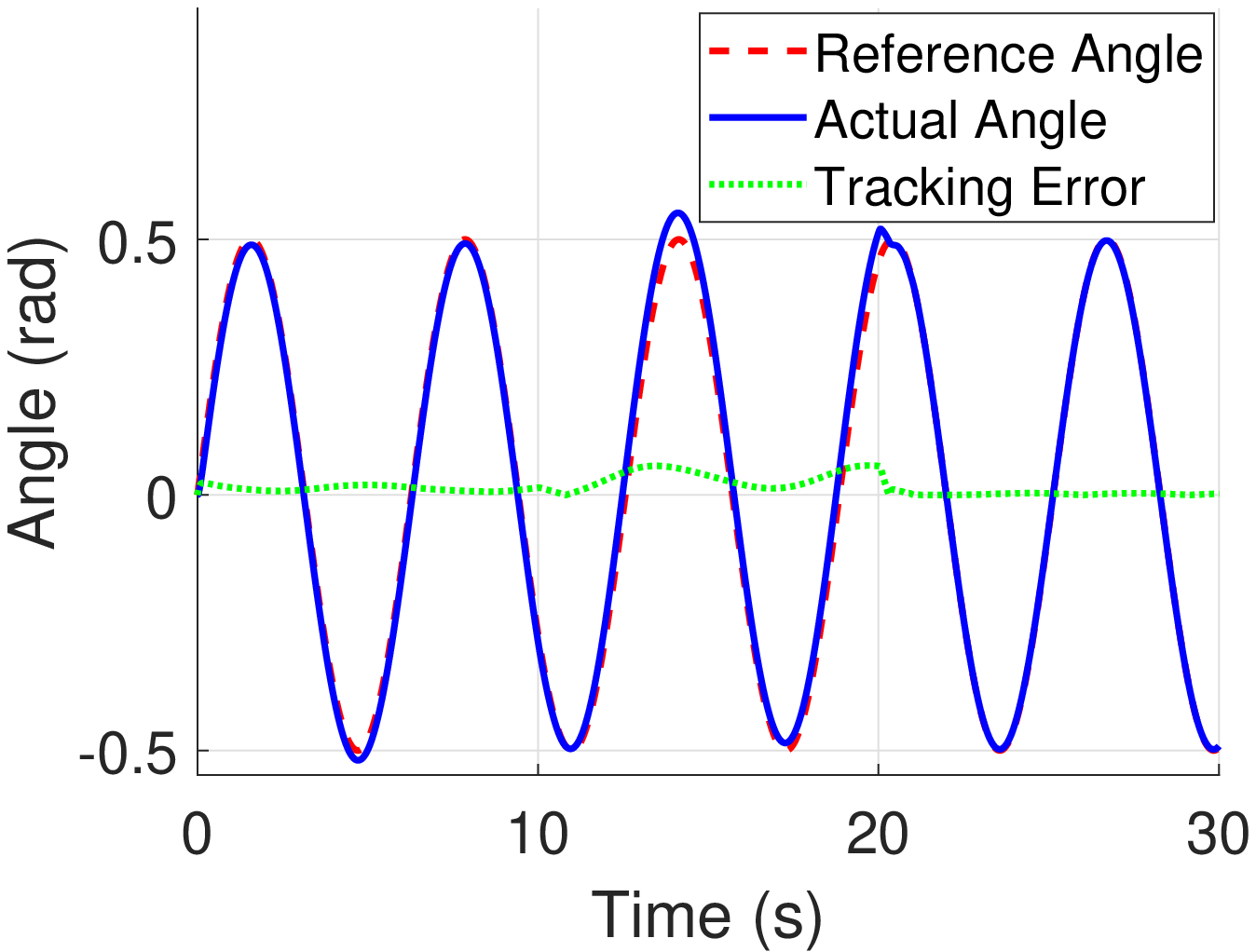}
\label{fig:incorrect_model_reference_signal_system}
}
\subfigure[Parameter estimation: Weights.]{
\includegraphics[width=0.64\columnwidth]{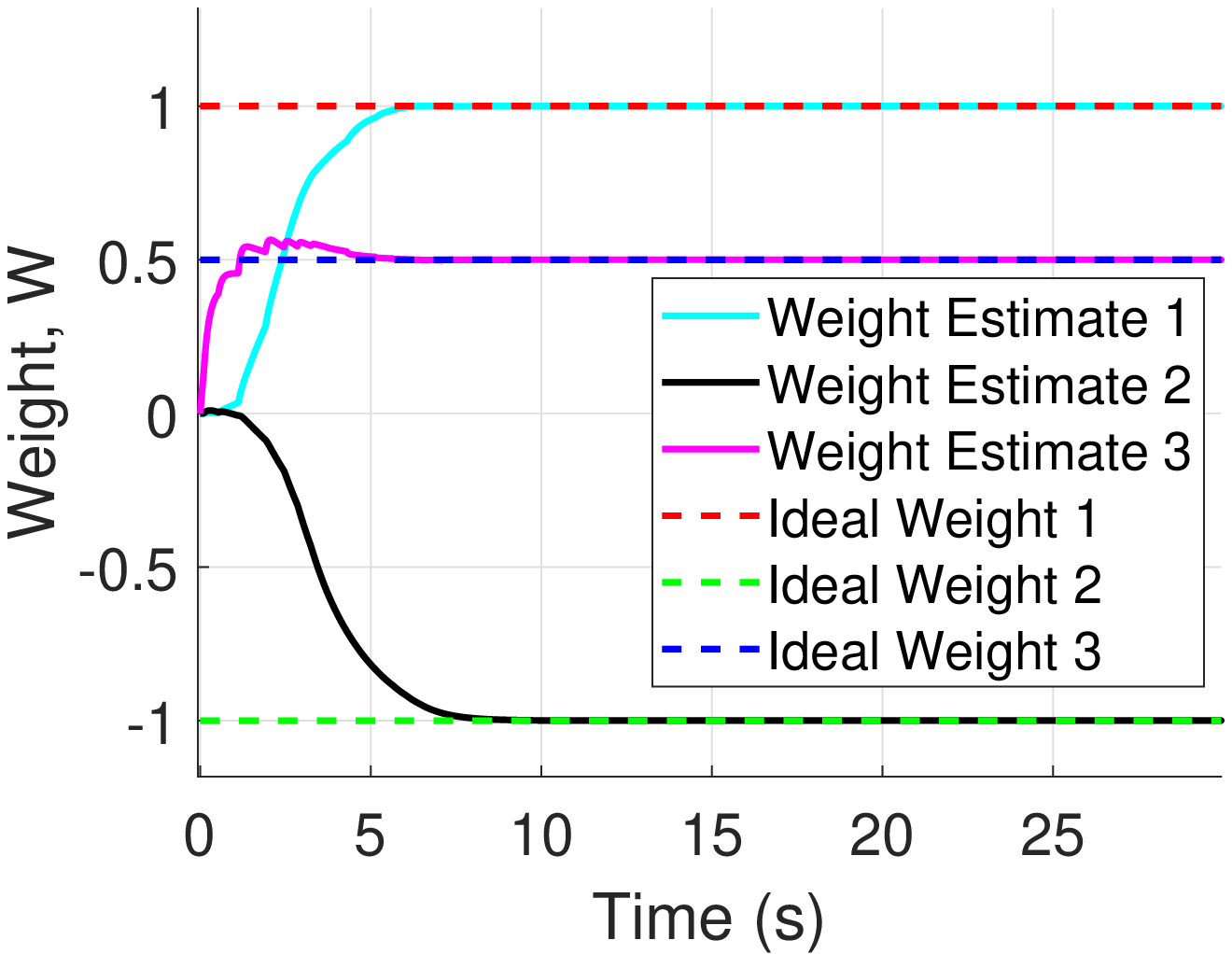}
\label{fig:parameter_weight_update}
}
\subfigure[Gaussian estimate of the disturbance.]{
\includegraphics[width=0.64\columnwidth]{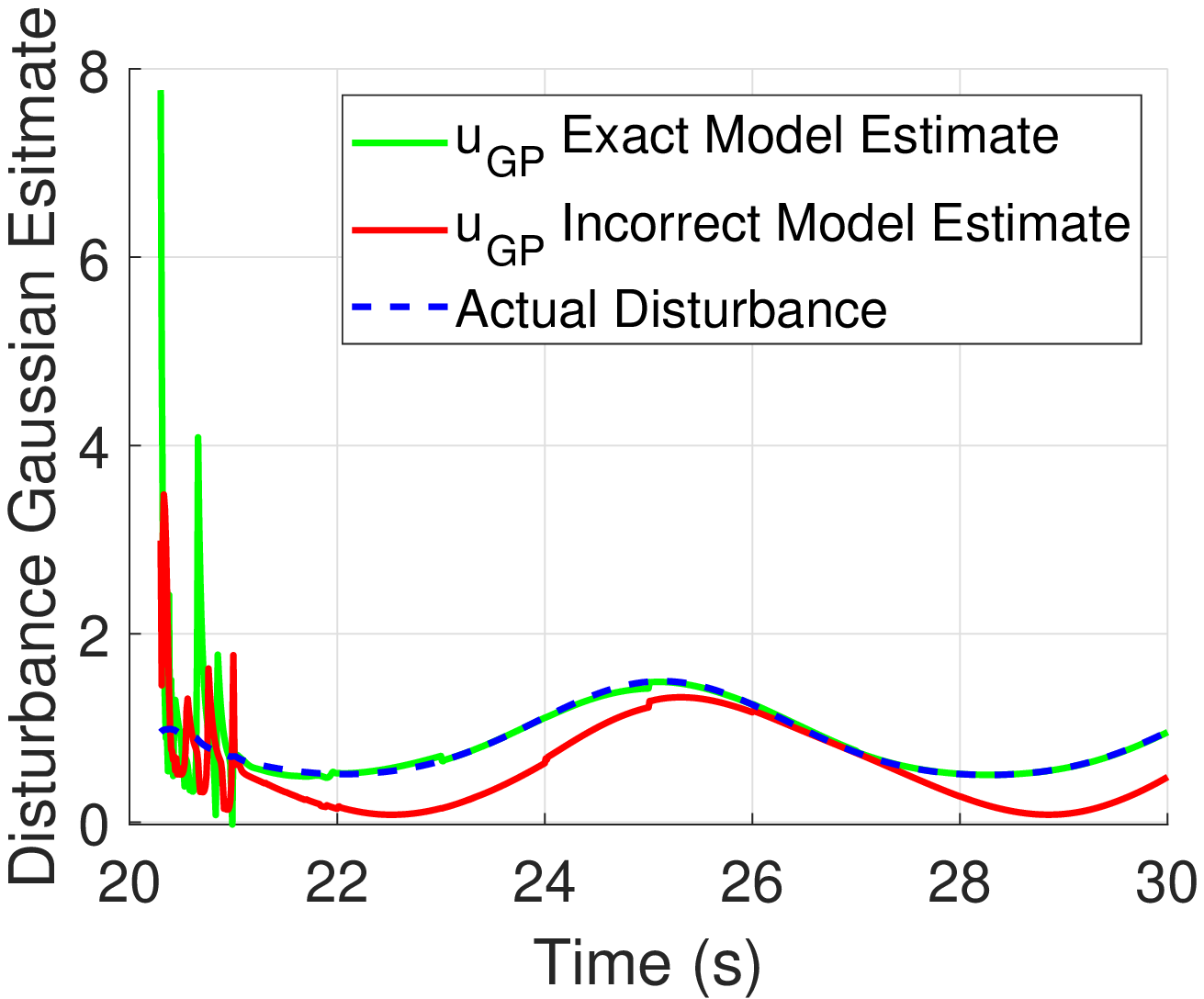}
\label{fig:Gaussian_estimate}
}
\subfigure[Output of the robustness term.]{
\includegraphics[width=0.64\columnwidth]{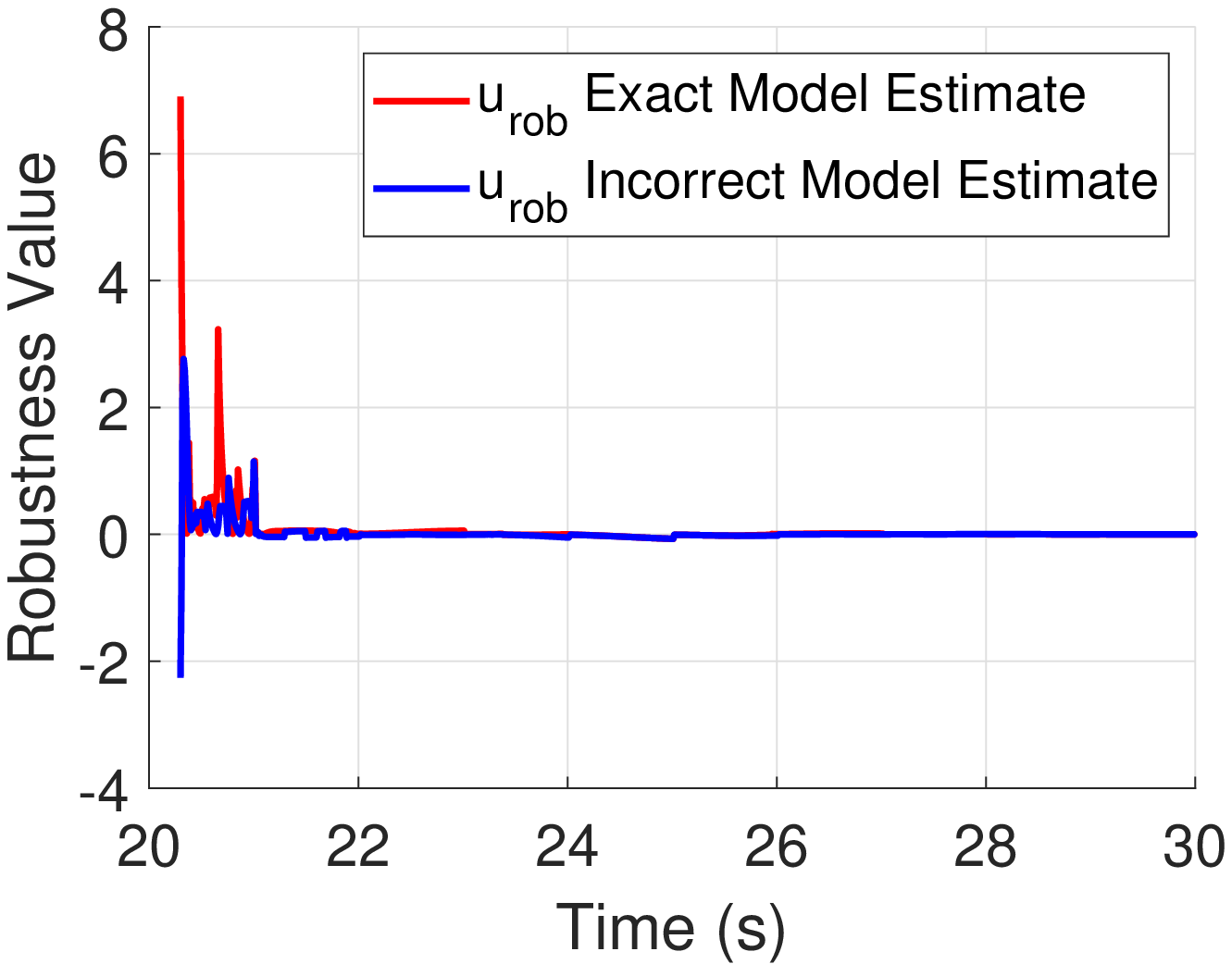}
\label{fig:robustenss_value}
}
\caption[Optional caption for list of figures]{Simulation results: \textbf{(a)} The average tracking error for each case outlined at the beginning of Section \ref{simulation}. Case d) and Case e) show the minimized tracking error when using the proposed control law. \textbf{(b)} Reference tracking when concurrent learning converges to the exact model parameters (\(\mathbf{\Tilde{w}} = 0\)) ( Case  e) ). 
        %The three stages in the simulation outlined in Section \ref{simualtion_systems} are shown. 
        \textbf{(c)} Reference tracking when concurrent learning does not converge to the true model parameters (\(\mathbf{\Tilde{w}} \neq 0\)) ( Case  d) ). \textbf{(d)} Exact parameter convergence using concurrent learning in Case b) and Case e). \textbf{(e)} Gaussian estimate of the disturbance when the model is exactly parameterized and when a model error is present ( Case d) and Case e) ). \textbf{(f)} The robustness value when the model is exactly parameterized and when a model error is present ( Case d) and Case e) ).}
\label{fig:}
\end{figure*}

The average tracking error for the cases outlined at the beginning of Section \ref{simulation} is shown in Figure \ref{fig:combined_average_tracking_error}. The proposed control law leads to the reduced tracking error in both scenarios, Case d) and Case e), even in the presence of large continuous disturbances. With the Gaussian Process learning and predicting the system disturbance, the average tracking errors for Case b) and Case e) are approximately equal. Both cases would have the same tracking error given that the GP exactly learns the disturbance. The effect of using a fixed model estimate versus a concurrent learning model is visible in Case a) and Case b), respectively, in which using concurrent learning drastically reduces the error. If a case was to exist where the model parameters were not estimated accurately using concurrent learning, the GP would learn the model error as part of the disturbance and provide appropriate compensation, as is shown by Case d). 

The three stages for Case e) and Case (d) are shown in Figure \ref{fig:correct_model_reference_signal_system} and \ref{fig:incorrect_model_reference_signal_system}, respectively. In the first stage, the model parameters are being learnt in a controlled environment for the first ten seconds only – this provides sufficient time for the parameters to converge to their parameter estimates. In the second stage, a disturbance is introduced, the GP also begins learning the disturbance as a torque mismatch for the 10 to 20 second period. During this time, the GP provides no compensation and only learns. After 20 seconds, the GP begins compensating for the effects of the disturbance and the tracking error is significantly reduced. The ability of the GP to compensate for large disturbances is apparent by comparing Case c) and Case e). In Case e), the Gaussian process captures the disturbance only as the model parameters are exactly estimated, whereas in Case d), the Gaussian process accounts for the disturbance and the system model error, as shown in Fig. \ref{fig:Gaussian_estimate}. The exact parameter convergence of concurrent learning in Case b) and Case e) is shown in Fig. \ref{fig:parameter_weight_update}. The robustness value for Case d) and Case e) is also shown in Fig. \ref{fig:robustenss_value}. As the GP compensates for the control input mismatch, the robustness term has minimal input. 

%%%%%%%%%%%%%%%%%%%%%%%%%%%%%%%%%%%%%%%%%%%%%%%%%%%%%%%%%%%%%%%%%%%%%%%%%%%%%%%%
\section{Conclusion}

This paper has developed the use of concurrent learning for exact model parameterization in the context of feedback linearization, combined with the use of Gaussian Processes for learning the disturbance for uncertain nonlinear systems. A robustness term was developed and added to the control law to ensure closed-loop system stability, which has been proven by using the Lyapunov stability theorem. Simulation results have verified the effectiveness of the control law in achieving minimized tracking error in the presence of model mismatch and disturbances, and ensured more accurate tracking performance.
%Beginning with the formulation of concurrent learning in the framework of feedback linearization, the error dynamics were formulated in the presence of an unknown disturbance. The Gaussian Process was used to model and then compensate for the disturbance in the proposed controller. To ensure closed-loop system stability, a robustness term was developed and added to the control law. The resulting system has been shown to be stable using the Lyapunov stability theorem. Simulation results have verified the effectiveness of the proposed control law in achieving minimized tracking error.  
% Previously, many papers have introduced the use of Gaussian-based learning, but the model parameter mismatch and disturbance cannot be distinguished. The proposed controller has the benefit of the providing exact parameter estimation in the context of feedback linearization in a controlled environment and accounting for disturbances using online Gaussian-based learning for disturbance compensation.
%\renewcommand*{\bibfont}{\normalsize}
%\printbibliography
\bibliography{ref.bib}

\begin{thebibliography}{10}
\providecommand{\url}[1]{#1}
\csname url@rmstyle\endcsname
\providecommand{\newblock}{\relax}
\providecommand{\bibinfo}[2]{#2}
\providecommand\BIBentrySTDinterwordspacing{\spaceskip=0pt\relax}
\providecommand\BIBentryALTinterwordstretchfactor{4}
\providecommand\BIBentryALTinterwordspacing{\spaceskip=\fontdimen2\font plus
\BIBentryALTinterwordstretchfactor\fontdimen3\font minus
  \fontdimen4\font\relax}
\providecommand\BIBforeignlanguage[2]{{%
\expandafter\ifx\csname l@#1\endcsname\relax
\typeout{** WARNING: IEEEtran.bst: No hyphenation pattern has been}%
\typeout{** loaded for the language `#1'. Using the pattern for}%
\typeout{** the default language instead.}%
\else
\language=\csname l@#1\endcsname
\fi
#2}}

\bibitem{westenbroek2020feedback}
T.~Westenbroek, D.~Fridovich-Keil, E.~Mazumdar, S.~Arora, V.~Prabhu, S.~S.
  Sastry, and C.~J. Tomlin, ``Feedback linearization for unknown systems via
  reinforcement learning,'' 2020.

\bibitem{8786140}
E.~{Kayacan}, ``Closed-loop error learning control for uncertain nonlinear
  systems with experimental validation on a mobile robot,'' \emph{IEEE/ASME
  Trans. Mechatronics}, vol.~24, no.~5, pp. 2397--2405, 2019.

\bibitem{8930275}
J.~{Umlauft} and S.~{Hirche}, ``Feedback linearization based on gaussian
  processes with event-triggered online learning,'' \emph{IEEE Trans. Autom.
  Control}, vol.~65, no.~10, pp. 4154--4169, 2020.

\bibitem{OLIVEIRA2020105927}
L.~Oliveira, A.~Bento, V.~J. Leite, and F.~Gomide, ``Evolving granular feedback
  linearization: Design, analysis, and applications,'' \emph{Applied Soft
  Computing}, vol.~86, p. 105927, 2020.

\bibitem{Erkansmlc}
E.~Kayacan, ``Sliding mode learning control of uncertain nonlinear systems with
  lyapunov stability analysis,'' \emph{Transactions of the Institute of
  Measurement and Control}, vol.~41, no.~6, pp. 1750--1760, 2019.

\bibitem{Slotine_Li_1991}
J.-J.~E. Slotine and W.~Li, \emph{Applied nonlinear control}.\hskip 1em plus
  0.5em minus 0.4em\relax Prentice-Hall, 1991.

\bibitem{Chai}
J.~Chai, E.~Medagoda, and E.~Kayacan, ``Adaptive and efficient model predictive
  control for booster reentry,'' \emph{Journal of Guidance, Control, and
  Dynamics}, vol.~43, no.~12, pp. 2372--2382, 2020.

\bibitem{9140024}
M.~{Greeff} and A.~P. {Schoellig}, ``Exploiting differential flatness for
  robust learning-based tracking control using gaussian processes,'' \emph{IEEE
  Control Systems Letters}, vol.~5, no.~4, pp. 1121--1126, 2021.

\bibitem{8264427}
T.~{Beckers}, J.~{Umlauft}, D.~{Kulic}, and S.~{Hirche}, ``Stable gaussian
  process based tracking control of lagrangian systems,'' in \emph{2017 IEEE
  56th Annual Conference on Decision and Control (CDC)}, 2017, pp. 5180--5185.

\bibitem{BECKERS2019390}
T.~Beckers, D.~Kulić, and S.~Hirche, ``Stable gaussian process based tracking
  control of euler–lagrange systems,'' \emph{Automatica}, vol. 103, pp.
  390--397, 2019.

\bibitem{5717148}
G.~{Chowdhary} and E.~{Johnson}, ``Concurrent learning for convergence in
  adaptive control without persistency of excitation,'' in \emph{49th IEEE
  Conference on Decision and Control (CDC)}, 2010, pp. 3674--3679.

\bibitem{8968461}
E.~{Kayacan}, S.~{Park}, C.~{Ratti}, and D.~{Rus}, ``Online system
  identification algorithm without persistent excitation for robotic systems:
  Application to reconfigurable autonomous vessels,'' in \emph{IEEE/RSJ Int.
  Conf. Intell Robots Syst.}, 2019, pp. 1840--1847.

\bibitem{Martins}
L.~Martins, C.~Cardeira, and P.~Oliveira, ``Feedback linearization with zero
  dynamics stabilization for quadrotor control,'' \emph{Journal of Intelligent
  \& Robotic Systems}, vol. 101, no.~1, p.~7, 2020.

\bibitem{MORENOVALENZUELA2020314}
J.~Moreno–Valenzuela, J.~Montoya–Cháirez, and V.~Santibáñez, ``Robust
  trajectory tracking control of an underactuated control moment gyroscope via
  neural network–based feedback linearization,'' \emph{Neurocomputing}, vol.
  403, pp. 314--324, 2020.

\bibitem{9102437}
G.~{Li}, A.~{Luo}, Z.~{He}, F.~j.~{Ma}, Y.~{Chen}, W.~{Wu}, Z.~{Zhu}, and J.~M.
  {Guerrero}, ``A dc hybrid active power filter and its nonlinear unified
  controller using feedback linearization,'' \emph{IEEE Trans. Ind. Electron.},
  pp. 1--1, 2020.

\bibitem{8944010}
M.~{Mehndiratta}, E.~{Kayacan}, M.~{Reyhanoglu}, and E.~{Kayacan}, ``Robust
  tracking control of aerial robots via a simple learning strategy-based
  feedback linearization,'' \emph{IEEE Access}, vol.~8, pp. 1653--1669, 2020.

\bibitem{Boley1994NumericalMF}
D.~Boley and B.~Datta, ``Numerical methods for linear control systems,'' 1994.

\bibitem{7858671}
R.~{Kamalapurkar}, B.~{Reish}, G.~{Chowdhary}, and W.~E. {Dixon}, ``Concurrent
  learning for parameter estimation using dynamic state-derivative
  estimators,'' \emph{IEEE Trans. Autom. Control}, vol.~62, no.~7, pp.
  3594--3601, 2017.

\bibitem{BRADFORD2020106844}
E.~Bradford, L.~Imsland, D.~Zhang, and E.~A. {del Rio Chanona}, ``Stochastic
  data-driven model predictive control using gaussian processes,''
  \emph{Computers \& Chemical Engineering}, vol. 139, p. 106844, 2020.

\bibitem{9285210}
Y.~{Cao}, J.~{Huang}, H.~{Ru}, W.~{Chen}, and C.~H. {Xiong}, ``A visual
  servo-based predictive control with echo state gaussian process for soft
  bending actuator,'' \emph{IEEE/ASME Trans. Mechatronics}, vol.~26, no.~1, pp.
  574--585, 2021.

\bibitem{Rasmussen_Williams_2006}
C.~E. Rasmussen and C.~K.~I. Williams, \emph{Gaussian processes for machine
  learning}, ser. Adaptive computation and machine learning.\hskip 1em plus
  0.5em minus 0.4em\relax MIT Press, 2006.

\bibitem{Duvenaud_2014}
D.~Duvenaud, ``Automatic model construction with gaussian processes,'' Ph.D.
  dissertation, 2014.

\bibitem{Steinwart_Christmann_2008}
I.~Steinwart and A.~Christmann, \emph{Support vector machines}, 1st~ed., ser.
  Information science and statistics.\hskip 1em plus 0.5em minus 0.4em\relax
  Springer, 2008.

\end{thebibliography}
\bibliographystyle{IEEEtran}
\clearpage

\end{document}